# Deep Layered Learning in MIR

Anders Elowsson

KTH Royal Institute of Technology
anderselowsson@gmail.com

**ABSTRACT**

Deep learning has boosted the performance of many music information retrieval (MIR) systems in recent years. Yet, the complex hierarchical arrangement of music makes end-to-end learning hard for some MIR tasks – a very deep and flexible processing chain is necessary to model some aspect of music audio. Representations involving tones, chords, and rhythm are fundamental building blocks of music. This paper discusses how these can be used as intermediate targets and priors in MIR to deal with structurally complex learning problems, with learning modules connected in a directed acyclic graph. It is suggested that this strategy for inference, referred to as *deep layered learning* (DLL), can help generalization by (1) – enforcing the validity and invariance of intermediate representations during processing, and by (2) – letting the inferred representations establish the musical organization to support higher-level invariant processing. A background to modular music processing is provided together with an overview of previous publications. Relevant concepts from information processing, such as pruning, skip connections, and performance supervision are reviewed within the context of DLL. A test is finally performed, showing how layered learning affects pitch tracking. It is indicated that especially offsets are easier to detect if guided by extracted framewise fundamental frequencies.

## 1. INTRODUCTION

Many of the skills necessary for successfully navigating modern human life are interrelated, but still so divergent that it is practically unfeasible to learn them all from one objective function. Although it may be possible to formulate an overarching goal for learning, progress generally stems from achieving smaller intermediate goals, gradually expanding a toolbox of concepts and procedures; acquired knowledge from one task can improve performance for other tasks (Ellis, 1965; Estes, 1970). For example, language acquisition presupposes contextual knowledge of the concepts encoded by each word (Lindsay & Norman, 2013). Many long-term projects in research and software development also evolve in a gradual fashion. New functionalities are developed on top of previously deployed and tested functionalities (Basil & Turner, 1975; Schwaber, 2004). Given the rise of machine learning for solving more and more problems in software development, it seems fruitful to discuss how interconnected learning architectures can be deployed for solving complicated tasks.

Music is a fitting environment for studying learning, an intricate arrangement of overlapping sounds across many dimensions. Pitches are layered across frequency to form harmony, and the pitched sounds are combined with percussion to form complicated rhythmical patterns. How can we teach machines to excel at music? Consider for example the complex task of writing and arranging music in a certain style using guidance from the audio files of a music album. Such a task requires the machine to be able to represent and understand music at a high level – something that necessitates several intermediate layers of non-linear transformations and representations. Many such intermediate representations are not perceptually well-defined, and to design these, human intuition only goes so far – decades of traditional research in MIR gave steady but rather slow incremental improvements. Deep learning has become an essential tool for bridging the unknown territory between input data and targets (Goodfellow, Bengio, & Courville, 2016; LeCun, Bengio, & Hinton, 2015), well suited for MIR (Humphrey, Bello, & LeCun, 2012). At the same time, some elements of music involving pitch, harmony, and rhythm, are well-established and fundamental to musical comprehension. Machine learning based systems may rely on such representations as intermediate targets to partition complex MIR tasks into two or more subtasks, thus breaking down complexity. Representations computed in one step may allow the system to reframe the subsequent learning problem, facilitating higher-level invariant processing. The learning steps necessary for untangling the musical components can be specified in a directed acyclic graph (DAG), using, for example, deep learning with musically appropriate tied weights at each individual step. This strategy for inference, referred to in this publication as deep layered learning (DLL), is starting to become prevalent in MIR systems with state-of-the-art performance (e.g., Elowsson, 2018; Krebs, Böck, Dorfer, & Widmer, 2016) and allows systems to account for the inherent organization of music.

The purpose of this publication is to provide an overview of DLL and to outline how such architectures can be designed to model music. A background to the rich machine learning repertoire that can be used is provided in Sections 2.1-3, while 2.4-5 reviews modular processing and layered learning in MIR. Section 3 offers a theoretical motivation and use cases focused on invariant processing of tonal representations. Section 4 describes relevant concepts of DLL and can be used as a guide for architectural design choices. Each concept is discussed in a dedicated subsection (4.1-7). Drawbacks with layered learning architectures are outlined in Section 5.1, providing some guidance as to when the strategy may *not* be successful. Section 5.2 discusses the potential of joint training strategies. In Section 6, a case study is done of a state-of-the-art polyphonic pitch tracking system that uses DLL, and Section 7 offers conclusions.

## 2. BACKGROUND

### 2.1 Representation Learning

A common challenge in machine learning is that available data are not organized in a way that facilitates inference. Classification will become much easier after the data is transformed into a new representation. For MIR, some suitable representations can be defined beforehand from knowledge about music and auditory perception. Researchers may compute a time-frequency representation (TFR) of the audio signal, and, as will be expanded upon in this paper, extract tonal or temporal information as intermediate representations. Commonly however, it is not known how to optimally represent the data at intermediate levels, so suitable representations (and transformations) must be discovered automatically (Bengio, Courville, & Vincent, 2013). Oftentimes, the best such *latent representations* (i.e., inferred and not observed) are derived after several layers of transformations. Depth generally leads to richer representations, yielding better results for MIR (Humphrey, Bello, et al., 2012).

Learning can be performed in an unsupervised way, by identifying compact representations from which the training set can be recreated (Hinton, Osindero, & Teh, 2006; Vincent, Larochelle, Bengio, & Manzagol, 2008). In MIR, this has been done for genre recognition (Hamel & Eck, 2010) and piano transcription (Nam, Ngiam, Lee, & Slaney, 2011) with a deep belief network (DBN). Supervised deep learning systems also learn abstract representation in hidden layers (Goodfellow et al., 2016; LeCun et al., 2015). In MIR, some of the earliest deep learning systems were designed to recognize genres (Li, Chan, & Chun, 2010) instruments (Humphrey, Glennon, & Bello, 2011), and chords (Humphrey, Cho, & Bello, 2012).

### 2.2 Transfer Learning

To successfully tackle new problems, previous knowledge from similar problems can be used (Perkins & Salomon, 1992). In machine learning, this induction from knowledge is called transfer learning (Pan & Yang, 2010), which can be used to manage a task when annotated data is scarce.

In MIR, several successful systems have first been trained on a large dataset with annotated tags, such as the *Million Song Dataset*, and then applied to smaller datasets for, e.g., genre classification (Choi, Fazekas, Sandler, & Cho, 2017; Hamel, Davies, Yoshii, & Goto, 2013; Hamel & Eck, 2010; Van Den Oord, Dieleman, & Schrauwen, 2014). Transfer learning has also been used for predicting musical preferences (Liang, Zhan, & Ellis, 2015) or playlist generation (Choi, Fazekas, & Sandler, 2016).

### 2.3 Architectures with Multiple Layered Classifiers

Many machine learning architectures have been proposed over the years involving multiple, layered classifiers, i.e., ensemble learning with a *dynamic structure* (Goodfellow et al., 2016). Early examples focused on dividing the input into nested sub-regions. Such algorithms, based on decision trees, include CART (Breiman, 1984), ID3 (Quinlan, 1986), and MARS (Friedman, 1991). The strategy was extended with neural networks (NNs) to perform the classification split (Guo & Gelfand, 1992), and separate gating networks to divide the input space for hierarchical mixtures of experts (Jordan & Jacobs, 1994). These systems have been described as modular NNs, but represent a different modularity than what is described in Sections 2.4-5 and Section 4.1, where a task is solved by subjecting all examples to *the same* learning modules (while these modules also can be used for other tasks).

A way to increase computation speed for classification is to use *cascading classifiers* (Alpaydin & Kaynak, 1998). With this methodology, weak classifiers are applied in a linear sequence to successively reject a larger and larger portion of the input space until (ideally) only true examples of the input space remain. A cascade of classifiers was used for face recognition back in 2001 (Viola & Jones). Another example is Google's system for address number transcription (Goodfellow, Bulatov, Ibarz, Arnoud, & Shet, 2014), where one classifier locates addresses and another transcribes them, as clarified by Goodfellow et al. (2016). Cascades have also been applied in MIR, to utilize the sparse distribution of pitched onsets for polyphonic transcription (vd Boogaart & Lienhart, 2009).

A related approach, *classifier chains* (Read, Pfahringer, Holmes, & Frank, 2009), is a method for transforming a multi-label classification task into a chain of binary ones. Labels are classified in consecutive stages, with the classification result at one stage supplied together with the original features for the next stage. Intermediate layers of the chain therefore act like hidden layers in a multilayer perceptron (Read & Hollmén, 2014), although each layer is trained with supervised learning. For comparison, this article will discuss ways to reframe complex tasks as multi-label classification problems, where intermediate targets preimpose the structural representation of the data. Classifier chains have been used in several systems for multi-label classification in MIR (Haggblade, Hong, & Kao; Read, Martino, & Luengo, 2013; Read et al., 2009).

Image segmentation and object classification has recently seen many models with intricate learning architectures (Zhao, Feng, Wu, & Yan, 2017), including: the part-based R-CNN (Zhang, Donahue, Girshick, & Darrell, 2014), convolutional neural networks (CNNs) combining bottom-up and top-down (Xiao et al., 2015) or locally shifting (through a recursive neural network, RNN) (Sermanet, Frome, & Real, 2014) attention mechanisms, and the CNN tree (Wang, Wang, & Wang, 2018) using several CNNs in a classifier chain. Notably, end-to-end DAG networks can use intermediate targets and average their gradients with those of the overall target during backpropagation, e.g., to guide structural segmentation (Flood, 2016). In robotics, neuro-controllers have been trained for several subtasks to perform a target task better; described, among other things, as modular decomposition (Doncieux & Mouret, 2014; Duarte, Oliveira, & Christensen, 2012; Urzelai, Floreano, Dorigo, & Colombetti, 1998).

### 2.4 Modular Music Processing

Many MIR systems have been designed in a modular fashion, using musical representations to express structural properties of the data at intermediate levels. There are several ways that such representations can interact. Framewise $f_0$ activations are commonly used for pitched onset detection, but they can also be used for beat tracking (Elowsson, 2016) and for computing a refined chromagram for chord detection (Mauch, 2010). Various perceptual features of music such as the "speed," rhythmic clarity, and harmonic complexity, can be predictive of music mood (Friberg, Schoonderwaldt, Hedblad, Fabiani, & Elowsson, 2014). The computed speed can be used together with periodicity estimates and tempo estimates for beat tracking (Elowsson, 2016). Beat estimates can be useful when trying to detect chords and chord changes (Mauch, 2010) aided by key estimation (Zenz & Rauber, 2007), or to detect downbeat positions (Durand, Bello, David, & Richard, 2015; Krebs et al., 2016). The opposite is also true; chord change information can be used when computing beat and downbeat positions (Goto & Muraoka, 1997; Papadopoulos & Peeters, 2008, 2011; Peeters & Papadopoulos, 2011; White & Pandiscio, 2015). Many systems have used harmonic-percussive source separation (HPSS) as an early processing step in an abundance of different tasks (e.g., Elowsson & Friberg, 2015; Gkiokas, Katsouros, Carayannis, & Stajylakis, 2012; Ono et al., 2010; Rump, Miyabe, Tsunoo, Ono, & Sagayama, 2010; Schmidt & Kim, 2013). This interdependence of musical concepts is probably why several MIR toolboxes (Eerola & Toiviainen, 2004; Lartillot, Toiviainen, & Eerola, 2008; McFee et al., 2015) have a modular design.

Modular music processing can be found when studying music impairments in brain-damaged patients (Peretz & Coltheart, 2003), leading researchers to suggest various modules for tonal and temporal perception. A biologically-inspired hierarchical unsupervised learning model has also been proposed (Pesek, Leonardis, & Marolt, 2014).

## 2.5 MIR implementations using layered learning

The modular music processing described in Section 2.4 can be extended to modular learning systems by using two or more learning steps and intermediate targets, factorizing the problem. Many successful systems using that approach have been proposed during recent years.

### 2.5.1 Pitch and melody

In polyphonic transcription, framewise $f_0$ estimation and note transcription have been separated into more than one learning algorithm in many systems. Marolt (2004) used networks of oscillators for partial tracking, and separate networks for detecting note pitches and repeated notes. Another early implementation (Poliner & Ellis, 2006) used a support vector machine (SVM) for the frame level classification, and a hidden Markov model with state priors determined from the training set, to identify onsets and offsets across time. A similar methodology was applied by Nam et al. (2011), using a DBN to extract features for classification. A musical model was then applied on top of the framewise predictions in a separate study (Boulanger-Lewandowski, Bengio, & Vincent, 2012), using an RNN combined with a restricted Boltzmann machine. In some implementations, the first step has been performed with unsupervised learning of latent variable models. Supervised classifiers have then been applied to refine the frame-level classification (Schramm & Benetos, 2017) or to perform note detection (Valero-Mas, Benetos, & Inesta, 2016; Weninger, Kirst, Schuller, & Bungartz, 2013).

A recent system jointly tracks pitch onsets and uses these for framewise estimation with CNNs (Hawthorne et al., 2017). The system is focused on piano transcription, and state-of-the-art results are reported for the MAPS dataset[i] concerning framewise predictions. An extended version adds a separate learning module for synthesizing the transcribed notes (Hawthorne et al., 2018). Manzelli, Thakkar, Siahkamari, and Kulis (2018) also applies a second module for synthesizing audio. Their input comes from an initial LSTM network trained to generate compositions. Pitch transcriptions can also be used as input to improve instrument recognition (Hung & Yang, 2018). Another recent system (Elowsson, 2018) first performs framewise $f_0$ estimation. Ridges of connected $f_0$s (corresponding to tone contours) are extracted, and onset/offset detection performed across the ridge (See Figure 1 in Section 3.1.1). Finally, tentative notes are extracted and classified. The system performs state-of-the-art across four datasets, MAPS[i], Bach10[ii], TRIOS[iii], and a Woodwind quintet[iv]. In Section 6, the effect of the separate learning steps in this system is analyzed further and compared to a direct onset and offset tracking.

### 2.5.2 Rhythm

Another set of MIR tasks where separate supervised learning steps can be useful is rhythm tracking (e.g., tempo estimation, beat tracking, and downbeat tracking). Some systems use one network for computing a time-varying activation curve, and then, to find the best sequence of beats, tune a few parameters for Viterbi decoding (Korzeniowski, Böck, & Widmer, 2014) or for a dynamic Bayesian network (Böck, Krebs, & Widmer, 2016). For downbeat tracking, the learning step for detecting downbeats has utilized beat synchronous input features, with the beats derived from a previous RNN learning layer (Krebs et al., 2016). A recent DJ system (Veire & De Bie, 2018) also uses beat-synchronous input features for downbeat tracking, and separate modules for solving other tasks (e.g., an SVM for singing voice detection). Elowsson and Friberg (2015) used linear regression to estimate the speed of the music and logistic regression to pick the tempo between several tentative tempi. Elowsson (2016) estimated the most salient periodicity with an NN and then used this to subsample periodicity invariant input features for a second network computing a beat activation. Speed and tempo were estimated using previously computed representations as input. The system performed state-of-the-art results on the Ballroom dataset.

## 3. THEORY, MOTIVATION AND USE CASES

### 3.1 Invariant Music Processing Facilitated by Intermediate Targets and Representations

The parameter sharing in deep learning is generally structured around a set of invariance or equivariance[v] assumptions about the input data. In image processing, a basic such assumption is spatial translation invariance, which spurred the development of the first CNN (LeCun et al., 1989). Relevant features generally have the same characteristics at different locations in the image, so CNN filter kernels can operate across the whole input space, producing *equivariant* filter outputs. By applying spatial dimensionality reduction, the *invariance* of the output features to irrelevant confounding factors, such as the viewing angle, is increased. Depth ultimately leads to progressively more invariant feature outputs (Goodfellow, Lee, Le, Saxe, & Ng, 2009), which improves the networks ability to make predictions. The importance of equivariant representations and parameter sharing for reducing the number of parameters and develop networks that generalize well is further described by Goodfellow et al. (2016: 329-335).

The spectrogram representation of music audio provides parallels to the spatial invariances of images, with time and frequency on the two axes. But many spatial invariances in music spectrograms are not locally constrained. Pitched tones extend sparsely across a large part of the spectrum and a log-frequency spectrum facilitates equivariant filtering with extended rectangular CNN filters operating across frequency (Thickstun, Harchaoui, Foster, & Kakade, 2018).

Layered learning strategies can instead be employed for dealing with, and promoting, more complex invariances of music. This includes invariant processing with regards to:

- Rhythm - including tempo, phase and metrical context (e.g., beat invariant processing as outlined in Section 2.5 to deal with tempo and phase).
- Tonality – including musical key and harmony.

Such strategies build on two assumptions in the context of invariant processing and intermediate representations:

(1) *Many musical representations are highly invariant.*

Many of the ways that musicians have chosen to represent music through the ages are highly invariant, stemming from the necessity to communicate musical ideas through concepts and symbols that can have the same meaning in many contexts. For example, music notation is invariant with regards to instrumentation, room characteristics, performed dynamics and associated spectral characteristics. Since the length and start of tones are encoded within a metrical structure, the representation is also invariant with regards to tempo; musicians can perform the sheet music at a tempo they find desirable. It can be argued that music notation emerged as the way to describe music *because* of the invariance it affords. With just a few black dots and lines on a piece of paper, musicians and composers were able to convey enough information to communicate the language of music, while its ultimate expression as the sound of a musical performance is very rich in variation and information.

We define the process of extracting musical representations as *structural disentanglement*, since these representations may come to span and relate to, e.g. the time-frequency space, in very complex non-linear ways. This leads us to the second assumption:

(2) *Structural disentanglement of music facilitates higher-level invariant processing.*

By extracting elements that define the musical organization (notes, beats, etc.), a new framework can be defined that facilitates higher-level invariant processing. In many cases, these invariances are expressed as a direct function of the musical organization, and their utilization (e.g., through parameter sharing) is therefore greatly simplified once this organization has been established.

The next sub-section provides examples of potential strategies involving both (1) and specifically (2) with regards to tonal representations. A further discussion of the general concepts is offered in Sections 4.2-3.

### 3.2 Use Cases Related to Tonal Representations

Extracted representations of music pitch can be used for reframing different learning problems. We will now review such strategies, starting with representations that commonly are predicted directly from spectrogram representations such as pitched onsets and frame-level $f_0$s, and then discuss progressively higher-level representations.

*3.2.1 At the tone-level*

To extract a musically meaningful structure from frame-level pitch estimates, pitch contours can be detected as ridges in "pitchogram" representations (e.g., Miron, Carabias-Orti, & Janer, 2014; Salamon & Gómez, 2012). Features of pitch contours can be used for, e.g., instrument recognition and genre detection, as outlined by Bittner, Salamon, Bosch, and Bello (2017). But these ridges can also form the basis of a new framework for subsequent networks, slightly transforming the input space. By extracting spectrogram bin energies at frequencies relative to the time-varying contour, a new input representation is formed that negates pitch fluctuations (.i.e., it is invariant with regards to them) (Elowsson, 2018). A network operating across this ridge (e.g., a CNN or an RNN) predicting onsets will operate both across time and pitch fluctuations at the same time. The functionality is visualized in Figure 1.

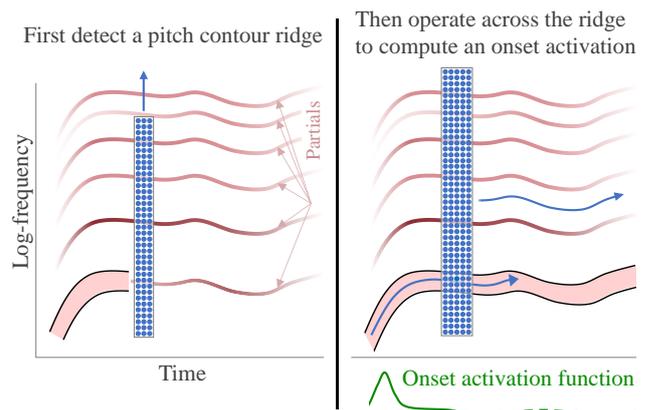

**Figure 1.** The process of first detecting a pitch contour ridge, and then operating across this ridge to facilitate tone-shift-invariant processing, e.g., for onset detection. The blue dotted rectangle represents a network kernel, and the blue arrows show the direction of operation. The pink broader ridge is a detected pitch contour.

Since the pitch fluctuations are negated when transforming the input space according to the pitch contours (e.g., a specific partial will occupy the same pitch bin for all time positions), networks that extend their processing across time can focus on more complex relationships between different time frames. For example, vocal contours have partial modulations based on sung formants and start and end points indicate a higher likelihood for fricatives. Related, the interrelationship between singing voice estimation and source separation have recently been utilized in several implementations (Chan et al., 2015; Ikemiya, Yoshii, & Itoyama, 2015; Stoller, Ewert, & Dixon, 2018).

Hawthorne et al. (2017; 2018) use onsets as a prior for $f_0$ (and offset) estimation in a joint training setup. The experimental analysis in Section 6 supports this idea regarding offsets; it shows that offsets can be more accurately predicted within the context of onsets and framewise $f_0$s.

### 3.2.2 Transcribed notes (not quantized)

At the next level, musical notes have been assigned an onset (and potentially an offset) but are not yet quantized into sheet music notation. There are many applications that can use this type of note representation; here we will focus on the task of aligning a predefined score with a music audio rendition. Audio-to-score alignment implementations have traditionally relied on matching templates generated from the score with corresponding features computed from the audio (Joder, Essid, & Richard, 2013), for example using the chroma. However, since the relevant level for alignment is the actual notes, there are strong reasons to first apply a polyphonic transcription system followed by an alignment algorithm, such as dynamic time warping (DTW). This can simplify the problem since it reduces the feature space across which matching is performed to a single variable. The activation of the matched feature should be a continuous probability estimate, allowing the system to output estimates with a higher recall. Recent systems are taking steps in this direction; Kwon, Jeong, and Nam (2017) used a polyphonic pitch tracker as an initial step before matching, showing that this can produce better results than approaches not relying on supervised learning to extract the musical structure. After alignment, additional learning algorithms can be applied, e.g., to adjust spuriously misplaced notes and to compute appropriate feedback to a student using the system for practice.

Transcribed notes have also been used in a layered learning setup for piano synthetization, where an initial system transcribes notes, and an additional system learns how to synthesize them (Hawthorne et al., 2018).

### 3.2.3 Music notation

At an even higher level of abstraction, the intermediate representation consists of data similar to sheet music notation, with transcribed note pitches quantized to the metrical structure. This representation has been used as a *starting point* for numerous machine learning systems, e.g., in the form of MIDI files, and it can be used to promote a broad range of invariances during processing including those outlined in the bullet-point list of Section 3.1. For example, by transposing all transcriptions according to a detected key, the processing of the subsequent learning module becomes invariant with regards to the key that the song was performed in.

One task for which music notation can be used as input is machine composition. Various relevant machine learning models have been proposed, including generative RNNs (Chu, Urtasun, & Fidler, 2016), long short-term memory (LSTM) networks (B. Sturm, Santos, Ben-Tal, & Korshunova, 2016) and CNNs (Yang, Chou, & Yang, 2017). In particular, the work by Manzelli et al. (2018) trained an LSTM network to generate compositions from MIDI and then used the generated compositions to condition a separate WaveNet (Van Den Oord et al., 2016) architecture rendering audio output.

Another example of a potential task at this level is the modeling of various subtle performance parameters related to the emotional expression (Juslin & Sloboda, 2001). Extracted music notation could provide a framework for analyzing differences between the tentative score and the performance (locating, e.g., phrase arches).

An abundance of other examples can be conceived by analyzing tasks that currently are performed on music notation input. Features to which the subsequent learning module has become invariant (e.g., tempo) can still be provided in various forms to provide context.

### 3.2.4 Instrument-specific models

Through instrument recognition, architectures with a dynamic structure can potentially be developed. A learning module is defined for each instrument (or instrument group when applicable), and this learning module is only used if an initial network has recognized that specific instrument. Such a design enables the system to solve tasks where the required processing differs wildly between different instruments, such as how the transcribed music should be played (e.g., appropriate fingering) and tasks related to the skill-level/complexity of a music performance. Parameter sharing can be designed based on instrument-specific invariances. For example, appropriate finger movements vary across black and white keys (Parncutt, Sloboda, Clarke, Raekallio, & Desain, 1997), and their relative position is repeated each octave on keyboards. Therefore, it can be appropriate to apply parameter sharing across octaves. For the initial instrument recognition, pitch-aware models can be used. Hung and Yang (2018) developed a system, where framewise pitch estimates from a separate CNN (Thickstun et al., 2018) improved performance.

## 4. IMPORTANT CONCEPTS IN DLL

As presented in Section 2.5, some recent systems for MIR use multiple learning steps, a few with a considerable depth in at least one step. As outlined in Section 3.2, there is an abundance of tasks where such architectures can be motivated from the perspective of acquiring higher-level representations that facilitate invariant processing. The architectures outlined in Section 2.5 and 3.2 are not generally motivated by the same circumstances as when multiple steps are used for transfer learning. The purpose is not only to transfer procedures acquired from one dataset to another, or to speed up performance by cascading the classification, but also to use several subtasks to distill knowledge of structural properties within intermediate representations. This is a method for approaching an overarching task with multiple annotations in a way that promotes generalization (see Sections 3.1 and 4.2) and, also, modularity in terms of the overall architecture (Section

4.1.2-4). As stated in the introduction, the approach will be referred to as *deep layered learning* (*DLL*), referencing a desirable depth of each learning step (i.e., deep learning), but also a potential depth of the layering of the learning steps.[vi] Several concepts important to DLL will be discussed in this section (here highlighted in italics with the corresponding subsection in parenthesis):

A DLL system contains *learning modules* (4.1) that infer a mapping between its input and a target with supervised learning. It is beneficial if intermediate targets have a high *validity* (4.2) with respect to the overall target. Invariant processing (see Section 3) can be promoted through *structural disentanglement* (4.3) of the music. Another learning module can then use the elements of the disentangled structure to form a framework for further processing while retaining their computed output activations and *latent representations* (4.4). *Skip connections* (4.5) from earlier points can also be employed to supply additional information as input to the next learning module. A layered architecture facilitates *pruning* (4.6) of the search space and can benefit from *layered performance supervision* (4.7) to tune precision/recall and identify weak links.

## 4.1 Learning Modules

We define a "learning module" as a module in which the relationship between input data and a target is inferred with machine learning. This subsection deals with concepts that are relevant when training each learning module separately. A discussion of joint training is then offered in Section 5.2. During run-time, a learning module processes its input data and supplies the output to the next processing step. The modules are arranged in a DAG together with other potential steps (e.g., DTW), producing higher-level representations at each hierarchical level. The predictive power of deep learning is best utilized if modules perform non-trivial tasks that require several layers of transformation. Learning is primarily broken down into modules to untangle the musical structure.

### 4.1.1 Training separate learning modules of a DLL system

A DLL system with separately trained modules can be trained in an iterative fashion by the following procedure: (1) process the training set (including the validation set) with any potential steps performed before a learning module to compute its input, (2) train the module, (3) process the training set with the newly trained module, (4) if there are more learning modules in the system, take the output and repeat from step (1). The architecture can have an overall training function for sequentially training and running the learning modules in an alternating fashion. If the input to a module is saved to disk in preparation of training (1), this data can also be used as a check-point so that subsequent processing of the training set can resume without reprocessing from scratch. However, the training set may vary between modules, in which case check-points cannot be used. That happens when the target annotations do not cover the same audio examples, or if regularization in the form of distortion (e.g., equalization, compression, added noise) is applied to the audio examples between the training of each module (Elowsson, 2018). Such techniques can make the system less reliant on particular output features from previous learning modules already fitted to the training set. Another technique for regularization (if the training set is very large) is to use different training tracks for training different modules. If a learning module relies on latent representations from a previous module, that input data will change drastically if the previous module is retrained[vii], in which case the later module also needs to be retrained. Output activations will not necessarily change as drastically, but subsequent modules can still be retrained to account for more subtle variations.

### 4.1.2 Modular intermediate targets

Figure 2 outlines important targets in MIR that can be used as intermediate representations for DLL, drawing on the examples in Sections 2.4-5 and 3.2. It is not to be interpreted as an exhaustive list of such targets but can be used as a starting point when designing new systems. Interrelated structures of pitch, rhythm, harmony and onsets are encapsulated by rounded rectangles (RRs), within which there is a strong interaction. The representations outside of the RRs are examples of slightly more high-level intermediate targets. Music notation refers to, e.g., quantized notes within a metrical structure and a predicted musical key, from which higher-level music analysis can be applied as outlined in Section 3.2.3; not the actual visual representations. All concepts in the figure have a rather clear perceptual and music-theoretical interpretation. The arrows in the figure indicate the direction in which learning modules can be connected. Note however that the encapsulated representations have all been predicted with deep learning in the past[viii]. The power of DLL rises when the representations in the figure are used as intermediate targets in more complex tasks, such as the examples provided in Sections 3.2.2-4 and 4.1.4.

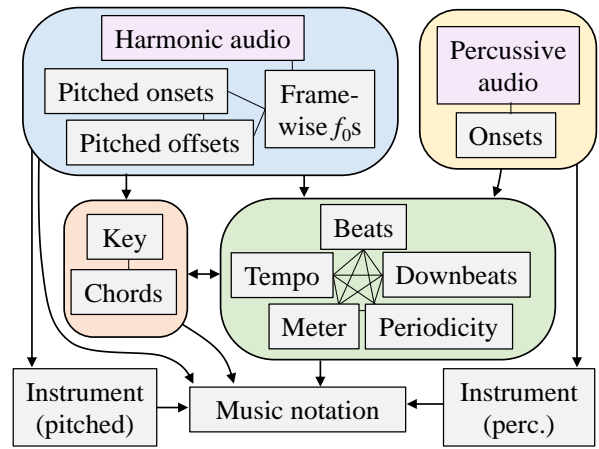

**Figure 2.** Important representations of music that can be used as intermediate targets in DLL. Additional learning modules can then be added in order to solve complex tasks with invariant processing. There is a strong interaction between rhythm representations (green RR), pitch-related representations (blue RR), the "vertical arrangement of pitch" (red RR), and percussive representations (yellow RR). Arrows indicate how different types of intermediate representations can be connected, though it should be emphasized that many representations can also be estimated directly from, e.g., a TFR. The output representations of HPSS are marked with purple.

Often in MIR, one type of annotations can be used to learn several relevant aspects of a task. Two such examples are shown in Figure 3. For polyphonic transcription, note annotations can be used to generate annotations for framewise predictions. For beat tracking, the annotated beat positions can be used for predicting a beat activation, the tempo, and to find the most likely sequence of beats. A system with several targets may therefore not necessarily need to rely on several distinct sets of annotations.

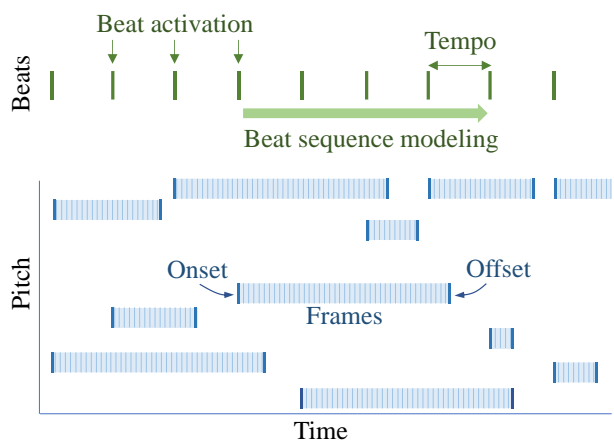

**Figure 3.** Illustration of how annotations for beat tracking (green) and polyphonic pitch tracking (blue) can be used to create multiple targets.

*4.1.3 Parallels to Modular Programming*

In computer programming, code modularity is an important design principle, allowing well-defined tasks to be abstracted into higher-level modules (Aberdour, 2007). For example, modularity during the development of Linux gave developers many benefits (Lee & Cole, 2000; Moon & Sproull, 2010) with parallels to those offered by DLL and learning modules.

**1)** Modular code is reusable since many modules can be used for other projects. In DLL, each learning module can be used for all tasks that depend on the representations computed by the module. Bittner et al. (2017) have highlighted this aspect to motivate pitch contour extraction.

**2)** When subtasks are divided into modules, it becomes easier to perform bug fixing, performance improvements, and other under-the-hood changes for each task separately. With DLL, it becomes possible to identify learning modules that are underperforming (Section 4.7). Performance can be maximized for each module, for instance, by experimenting with different parameter sharing mechanisms.

**3)** Modular code can be developed by large teams, as team members can focus on separate code parts that interact only through pre-defined interfaces. Modular DLL systems can be developed by several research groups (e.g., Hung & Yang, 2018), as modules can interact through pre-defined representations.

*4.1.4 Role in large-scale intelligent systems*

Systems that rely on machine learning for deduction may still require substantial development efforts. Input data often needs to be pre-processed, preparing training data can be costly and time-consuming, and a suitable structure for, e.g., parameter sharing needs to be determined in an experimental fashion. Time-to-market or time-to-publication can conceivably be reduced if a model of relevant sub-targets already exists when a new opportunity emerges. The cost of developing high-performing models motivates the reuse of these models when trying to predict new targets for which the old model computes a relevant sub-target. No matter the depth of a model, there are often tasks where its output predictions and latent representations can form the basis for pursuing higher-level targets.

An example pertaining to the role of learning modules: The output of a trained polyphonic pitch tracker can be used as input when creating a complete music transcription system. The music transcription output can be used by an additional module to analyze students' compositions and give feedback in an interactive app. Data from student interactions are unlikely to be useful for improving pitch tracking capabilities through end-to-end learning, but well suited for adjusting how predictions at various levels of the system should be interpreted.

**4.2 Validity**

Deep layered learning can be used to force the system into using certain intermediate representations or processing steps that the researcher knows are important – to enforce validity. This can be useful for tasks composed of rather independent subtasks (Gülçehre & Bengio, 2016), and also prevent overfitting, as the system cannot to the same extent make predictions by inferring complex and irrelevant relationships between the input data and the target. Irrelevant in this context refers to those relationships that will not generalize outside of the training set as discussed by Sturm (2013, 2014) in the context of genre recognition.

Consider an app that generates chord progressions to vocal performances sung across drum beats. The app developer knows that beats, downbeats and rhythmical accents signify time positions where chords changes are common; the relationship between chord changes and downbeats has been utilized in the past (Papadopoulos & Peeters, 2008, 2011; White & Pandiscio, 2015). The system can rely on a rhythm tracker for computing probable position for chord changes and a pitch tracker trained to extract vocal melodies sung across beats. A final learning module utilizing these representations for computing chords will not to the same extent rely on assumptions about how, e.g., vocal timbre, background noise or rhythm sounds covaries with annotated chords. Such relationships may exist only in the training set and are therefore less "valid" for the task. With a DLL setup, the developer also has more direct measures for controlling the extent to which certain covariations in the training set are allowed to influence the system. Drum sounds may correlate with chord complexity, e.g., if the training set contains examples of jazz. If spectral characteristics of the drums are supplied to the final chord progression module, the developers imply that such correlations are valid.

A parallel can be made to structured probabilistic models that convey information by leaving edges out, thereby specifying an assumption that any interaction through that edge is irrelevant (Goodfellow et al., 2016: 579).

**4.3 Structural Disentanglement**

Jo and Bengio (2017: 1) outline two ways for machine learning models to achieve good generalization: capture high-level abstractions, or overfit to superficial cues present in both the training and test set; the latter option implicates low validity as touched upon by Sturm in the previous subsection. The propensity for image classifiers to fail on adversarial examples (Szegedy et al., 2013), and surface statistical regularities indicating that they overfit to superficial cues, has motivated research into how to design networks that extract more abstract disentangled representations at intermediate layers of processing (Jo & Bengio, 2017). A step forward seems to be to build in invariance through prior domain knowledge (Bengio et al., 2013).

We argue that a prior domain knowledge appropriate for music is that of the inherent musical organization. A MIR system can use intermediate targets of the musical representations at some steps to constrain processing. These intermediate representations are extracted based on computed activations so that a musically organized representation emerges. This disentanglement may vary in complexity and scope. It can involve thresholding of a single activation to derive a note onset, or full notation of polyphonic a capella music accounting for both tempo and pitch drifts. We refer to this process, in general, as structural disentanglement. The appropriate methodology for performing structural disentanglement depends on the task at hand. Section 4.1.1 outlined the general training process when the architecture is partitioned into separate learning modules at the disentanglement step, but some tasks allow for gradient-based solutions facilitating joint training (Section 5.2). The extracted structural representations can then facilitate higher-level invariant processing as previously outlined (Section 3).

### 4.4 Latent Representations

The latent representations in the last hidden layer of a learning module consist of various features useful for predicting the target of the module. When computing a framewise beat activation (see Section 2.6.2), one neuron in a hidden layer may activate if a kick drum is present, another neuron for the snare drum, and some neurons may activate if periodical musical accents intersect the present frame. Many neurons will also activate to attributes for which no clear musical terminology exist. It is reasonable to assume that these hidden layer activations can be useful, in addition to the actual output prediction, for predicting the target of a subsequent learning module – after all, a major point of DLL is to extract high-level music representations. Therefore, it may be beneficial to use representations from the last hidden layer of the earlier learning module when predicting a new higher-level target. In this respect, the processing can be compared to the layerwise training of DBNs (Hinton et al., 2006; Nam et al., 2011).

### 4.5 Skip Connections

Skip connections were introduced more than 20 years ago for feedforward networks (Kalman & Kwasny, 1997). They allow lower-level representation to skip layers of processing in the network. Recently, skip layers or "residual layers" have also been proposed for deep learning with CNNs (He, Zhang, Ren, & Sun, 2016). In MIR, a U-net architecture with skip connections has been used for singing voice separation (Jansson et al., 2017).

There are many situations where it can be necessary to design systems such that a learning module *B* receives input from previous stages via skip connections, covering for aspects of the data not captured by an immediately prior module *A*.[ix] This can happen when *A* was computed to:

- extract the musical structure to facilitate higher-level invariant processing (see Section 3 and Figure 1), or
- allow for complex intermediate processing chains where it is hard to track a gradient (e.g., DTW).

### 4.6 Pruning

Intermediate targets can often be used to prune the search space during run-time. This may be beneficial in MIR since many musical elements such as tones, beats and onsets are sparsely distributed across the input space (e.g., a TFR). Thus, if the first learning module in a DLL system identifies sparse musical elements, the overall system can function like cascading classifiers (Section 2.3), speeding up processing. Polyphonic transcription is one task where pruning has been applied to speed up computation (Elowsson, 2018; vd Boogaart & Lienhart, 2009).

### 4.7 Layered Performance Supervision

The training annotations of a DLL system can be used to evaluate each module separately, providing insights into strengths and weaknesses of various steps (parallels to modular coding were provided in Section 3.1.3). But the evaluation of performance at each intermediate target can *also* provide useful information with respect to the overall target. Such evaluations may be used to define an upper limit on performance for the overall target incurred by an intermediate learning module. For example:

- When performing pitched onset detection from framewise $f_0$ ridges, the proportion of annotated onsets that are close to any activations of framewise $f_0$s defines the upper limit.
- When performing downbeat tracking from input features synchronized to faster pulse levels, the proportion of annotated downbeats that coincide with this level defines the upper limit (Durand et al., 2015).

In DLL systems where structural elements of the music are extracted, it will be beneficial to tune the balance between precision (the proportion of detected the targets that were correct) and recall (the proportion of the annotated targets that were detected) based on the evaluated intermediate performance and the overall upper limit. For example, by retaining estimated $f_0$ frames that have a lower probability than 50 % to be correct, longer contours will not be disrupted by spurious missed frames. Subsequent learning modules can always dismiss such frames later, based on their activations, if that makes sense in a wider context. The merit of having a high recall of pitch contours was discussed by Bittner et al. (2017). Higher recall may also have negative effects:

- If there is a point at a higher recall where the musical organization becomes less clear. In the example with $f_0$ contours, false $f_0$ activations may incorrectly connect contours into longer nested tones.
- If computation speed is important. Higher recall reduces pruning capabilities, especially if the extracted structures are computationally expensive to process.

# 5. SHORTCOMINGS AND JOINT TRAINING

## 5.1 Shortcomings

We will now review four risk factors that are important to be aware of when designing DLL systems:

(A) *Propagation of inaccuracies* – the risk that inaccuracies of a learning module propagates in such a way that the system cannot accurately predict the end target. To compensate for this, previously discarded information can be provided through latent representations (Section 4.4) and skip connections (Section 4.5). But whenever systems trained with separate learning modules relies on a module for a structural understanding, e.g., to extract music notation (Section 3.2.3), tone contours or input for music alignment matching (Section 3.2.2), the performance of that module becomes a critical factor for success. In connection to this: processing steps performed outside of any learning module may also introduce inaccuracies, which then propagate further in the system.

(B) *Complex architectures that become hard to train and maintain* – relying on several learning modules trained separately may make training more time consuming, and requires that an overall function is developed for automating the process. Furthermore, the architecture may require later modules in the DAG to be retrained following an update of an earlier module according to the conditions outlined in Section 4.1.1. Complex architectures that rely on a larger code base and several sets of annotations will become harder to maintain, and the code harder to port to a new coding language.

(C) *Inaccurate assumptions about optimal intermediate representations* – suboptimal intermediate targets may lead to architectures where extracted musical representations are not invariant with regards to expected musical variations, or do not support invariant processing in higher-level learning modules (see both points in Section 3.1). This risk factor includes intermediate targets with low validity (see Section 4.2), which is especially critical if DLL is used mainly for transfer learning. Note that assumptions are part of all machine learning driven systems to varying degrees; here we only focus on assumptions related to intermediate representations.

(D) *Training data bottlenecks* – the system may come to rely on a learning module for which it is hard to develop new training data. With too few training examples at intermediate targets in relation to what is available for the overall target, benefits incurred by transfer learning will not apply. Problems may emerge over time in systems that gradually expands the training set for the overall target, for example by tracking users' interaction with the system. In the example provided in Section 4.2, the chord generating app may after a few months have collected a huge library of user-supplied chord progressions, and therefore generalize better if these are used directly in end-to-end learning.

## 5.2 Joint training

As described in Section 2.3, end-to-end DAG networks can use intermediate targets, guiding the learning (e.g., (Flood, 2016; Hawthorne et al., 2017). In many scenarios it can be good to use such joint training methods, instead of separate learning modules in MIR. An architecture that does not rely on complex steps between learning modules can pass a gradient from the higher-level network to the lower in the DAG during training, thereby combining networks into a larger. Such techniques can be useful for enforcing validity (Section 4.2) or for pruning the search space (Section 4.7) of the second network. They are however in many cases harder to use for achieving higher-level invariant processing as outlined in (2) of Section 3.1, whenever this requires a restructuring of the input data at intermediate steps. Gradient-based solutions for performing structural disentanglement (Section 4.3) in MIR is therefore an important question for future research.

In the context of validity, it could be interesting to explore weighting schemes when combining gradients at hidden layers during backpropagation (other than taking the mean). With high weights for the gradients of intermediate targets, the DAG network will function more like the DLL systems described in this paper, or as the DBN used for chord recognition by Boulanger-Lewandowski, Bengio, and Vincent (2013); with low weights, it will perform more conventional end-to-end learning.

It should be noted that there is no reason for why a joint training network cannot be combined with a separate learning module for solving complex tasks. One such example is the system by (Hawthorne et al., 2018) for piano music modeling *and* generation. As previously outlined, the appropriate training methodology depends on a number of factors including the complexity of the structural disentanglement, the validity of the intermediate targets, and the available training data.

# 6. THE EFFECT OF DLL ON POLYPHONIC PITCH TRACKING

This section evaluates various aspects of the layered learning architecture of a state-of-the-art polyphonic pitch tracking system described by Elowsson (2018). This system first computes pitch contours with a high resolution, then detects onsets across these contours, and finally evaluates each detected onset in relation to neighboring onsets. The F-measure ($\mathcal{F}$), which is the harmonic mean of recall and precision, was computed at *three relevant steps* of learning for framewise pitch estimation ($\mathcal{F}_{fr}$), pitched note onset tracking ($\mathcal{F}_{on}$), and pitched note offset tracking ($\mathcal{F}_{off}$). By trying to maximize performance at each step, it became possible to evaluate how much performance increases throughout the learning modules. Onset tracking was evaluated allowing deviations of 50 ms, and offset tracking was evaluated allowing deviations of 100 ms; a limit of 50 cents was also used in all cases. Furthermore, a *new* polyphonic pitch tracking system was implemented trying to predict pitched onsets and offsets directly, using the same filtered spectrogram input as the original system. All variations of the systems were trained on the training set developed by Elowsson (2018) and then tested on four different test sets, Bach10[ii], MAPS[i], TRIOS[iii], and the Woodwind quintet[iv]. The evaluation metric was finally computed as the harmonic mean of the results for the four separate test sets, in accordance with the methodology proposed by Elowsson (2018).

## 6.1 Evaluation Design for the Original System

The evaluation was performed at three steps (1-3) of the layered learning architecture:

1. The system operates across pitch to perform framewise $f_0$ estimation, rendering a Pitchogram representation (a piano-roll with centitone resolution). At this point, $\mathcal{F}_{fr}$ was estimated, with a threshold optimized on the training set. Pitch contours of tones will appear as ridges in this representation, and these ridges were extracted by thresholding. *Two* additional thresholds were then determined through a grid search on the training set, selecting the thresholds that maximized $\mathcal{F}_{on}$. First, the pitch ridges were thresholded based on summed activations across the ridge. Pitch ridges below the determined threshold were removed. Then, an onset position was determined as the *first* time frame of the ridge with an activation above the second threshold. The offset position was instead determined as the *last* time frame above a slightly lower threshold.

2. Onset and offset activation networks are then employed by the system. These networks use the ridges to form a new framework across which they operate, as illustrated in Figure 1 and explained in the text (Section 3.2.1). At the second point of evaluation, onsets were estimated after peak-picking the computed onset activation. Offsets were extracted as the first time-frame above a threshold (0.5) applied to the activation of a network trained to output 1 at all positions beyond the note offset in a ridge, and 0 at all positions before the offset. For onsets, the same four parameters for filtering the activation curve described by Elowsson (2018, Section 8.6) were determined in a grid search on the training set, but the weighting function in Eq. 20 of that article was not used, opting instead to simply take the combination of parameters that maximized $\mathcal{F}_{on}$. The same methodology as the one proposed by Elowsson (2018, Section 9) was used for thresholding offsets, once again optimized on the training set. Based on onsets and offsets, tentative notes were derived, and framewise estimates were computed along the contour of these notes.

3. After tentative notes have been analyzed in the last network, and incorrect notes removed, the final estimates were computed. Once again, the start and end of notes defined onsets and offsets, and framewise estimates were computed for pitched frames along the ridges between note starts and note ends. For this step, no new processing was introduced, and results are therefore reported according to Elowsson (2018).

The thresholds used to compute intermediate performance in step 1 and 2 were not applied at later steps. Instead, a higher recall (motivated in Section 3.7) was used by the intermediate modules (as in the original publication).

### 6.2 Design of the New Comparison System

A new system was designed that tries to estimate onsets and offsets directly, to provide a comparison with the layered system. The new system used the same filtered spectrogram and spectral flux (SF) input computed from the variable-Q transform (Schörkhuber, Klapuri, Holighaus, & Dörfler, 2014) that the original system received through skip-connections (while operating across the ridge). This input is a rectangular kernel (size 247x4 for the SF) ranging between −93 to +153 bins relative to each evaluated frequency with frequency bins spaced 40 cents apart, and ranging between −46.4 ms to 23.2 ms relative to each evaluated time-point with bins spaced 23.2 ms apart. The filtered spectrogram input was also provided, just like in the original system, at each evaluated time position across the same frequency range. Furthermore, the same vector covering loudness variation from the initial filtering was provided, as well as the pitch. The new system also used the same network size for the hidden layers, the same training method and the same training set.

By necessity, the post-processing differed between the two systems, since the new system could not rely on the pitch contour to define a one-dimensional framework in which onsets and offsets were detected by peak-picking (See Figure 1). Therefore, to give the new system the ability to also handle fluctuating pitches from vibrato at onset positions, it was designed to track onsets in two-dimensional space. Care was taken to emulate a similar behavior across two-dimensional space, and to develop an architecture that could produce competitive results. The processing was done on the two-dimensional activation matrices $X$ of onset and offset activations computed for all pitches between MIDI pitch 26 and 104, spaced 20 cents apart, with a hop size of 5.8 ms. As in the original system, the activations were collected prior to the final sigmoid activation function of the networks. First, a smooth threshold $t$ was applied to all bins of $X$, in accordance with Eq. 17-19 of the original publication (Elowsson, 2018). Then, $X$ was smoothed with a two-dimensional Gaussian filter $f$, and a new threshold $t_2$ applied. After filtering and thresholding, onsets and offsets will appear as small "blobs" or "regions" in $X$, and these were extracted, just as pitch ridges were extracted in the original system. The thresholding parameters and the width of the filter that maximized $\mathcal{F}_{on}$ and $\mathcal{F}_{off}$ were determined in a grid search on the validation set, where each parameter could be at a reasonable low or high setting. The filter $f$ was σ = 1.8 across frequency and σ = 0.85 across time. The threshold $t$ was −8 for both onsets and offsets, whereas the optimal threshold $t_2$ was 1.5 for onsets and 3 for offsets. The location of the maximum activation peak within each detected region defined the position of the tentative onset or offset, and the value of this peak defined a reliability. Just as in the original system, an exhaustive search was performed on the validation set to determine the reliability value threshold that maximized $\mathcal{F}$, and this value was used to make a final decision.

### 6.3 Results and Discussion

Table 1 shows the combined harmonic mean results at three different steps of processing in the DLL system, and the results for the new direct pitch tracking system.

For the DLL system, performance increases throughout learning, with the most significant increase at the first learning module dedicated to the specific subtask (step 1 for $\mathcal{F}_{fr}$, and step 2 for $\mathcal{F}_{on}$ and $\mathcal{F}_{off}$). The direct system had an F-measure that was 3.9 lower for onset detection and 6.8 lower for offset detection. It seems reasonable that performance drops more for offset detection than onset detection since onsets oftentimes also are more directly perceivable in music. Offsets can in many cases to a larger extent be determined through a more elaborate analysis of the music structure (both for listeners and, apparently, ma-chine learning systems). For example, offsets of plucked or hammered instruments, such as piano, can more easily be determined through knowledge of a preexisting onset or by tracking a slowly decaying $f_0$ over time.

|  | **Step** | $\mathcal{F}_{fr}$ | $\mathcal{F}_{on}$ | $\mathcal{F}_{off}$ |
|---|---|---|---|---|
| **DLL** | 1 | 75.1 | 38.1 | 32.0 |
|  | 2 | 76.7 | 83.1 | 64.3 |
|  | 3 | 78.1 | 85.9 | 66.6 |
| **Direct** | - | - | 82.0 | 59.8 |

**Table 1.** F-measure ($\mathcal{F}$) at three different points of processing (step 1-3) in a DLL architecture for polyphonic pitch tracking, compared to the results of a new system that estimates onsets and offsets directly. Results are shown for framewise estimates ($\mathcal{F}_{fr}$), onset detection ($\mathcal{F}_{on}$), and offset detection ($\mathcal{F}_{off}$) as the harmonic mean results from the four test sets.

Although the original system computes onsets already at step 2, there are two reasons for comparing the direct results with the results of step 3: (1) The staged thresholding and blob detection of the new direct system was designed to handle uncertainties concerning onsets of sustained instruments with a slower attack since the blob will form a larger continuous region. These onsets can sometimes produce two onsets in the DLL system (step 2), which then are corrected in the final network that analyzes context (step 3). (2) An important point of the layered architecture is to extract structural elements (onset, offset and contour) for further processing, which is done thoroughly at step 3. For an extensive comparison with other methodologies for polyphonic transcription, including deep learning, the reader is referred to the previously cited publication (Elowsson, 2018). Both analyzed systems compare favorably to a baseline system (Dressler, 2017) evaluated to $\mathcal{F}_{on}$ = 59.1 and $\mathcal{F}_{off}$ = 45.0 in that publication.

## 7. SUMMARY

This article has discussed how architectures can be trained with intermediate targets to account for the inherent organization of music, partitioning complex MIR tasks into two or more subtasks. As shown, a growing number of publications make use of such strategies for music processing, and there are many potential use cases for future systems. The role of intermediate targets for facilitating invariant music processing and parameter sharing was outlined in Section 3. By unveiling the musical organization in the first learning step, it becomes possible to restructure the data such that additional learning steps are invariant with regards to variations (for example key or tempo) that are irrelevant to a particular task. Section 4 discussed the training procedure of separate learning modules and the role of skip connections for retaining information after the structural disentanglement. This section also highlighted how intermediate targets can be used to prune the search space and evaluate intermediate processing steps. Risk factors associated with DLL were outlined in Section 5.1, including the propagation of inaccuracies in potentially complex architectures that become hard to train and maintain. But as discussed in Section 4.1.3-4, the methodology may also offer a way to use existing fully functional learning systems as sub-models when predicting higher-level targets, reducing development time and cost. Joint training strategies were discussed in Section 5.2. Such strategies are a viable option whenever a gradient-based solution for performing the structural disentanglement exists. Experimental results presented in Section 6 for polyphonic pitch tracking shows how performance increase throughout learning, and indicates that structurally-aware models achieves better performance than direct ones for the investigated task.

---

[i] http://www.tsi.telecom-paristech.fr/aao/en/2010/07/08/

[ii] http://www2.ece.rochester.edu/~zduan/resource/Bach10%20Dataset_v1.0.pdf

[iii] https://c4dm.eecs.qmul.ac.uk/rdr/handle/123456789/27

[iv] https://docs.google.com/forms/d/e/1FAIpQLSd0ZnoJr4V-oiIO-zOg0jniNLBlwgKqP_NnY2oGVyLYqkXihRg/viewform?c=0&w=1

[v] We will use the term "invariant" to describe the many cases where both applies, i.e. "invariant or equivariant;" and use the term "equivariant" in cases where that term applies (more or less) exclusively.

[vi] I.e. both "deep-layered learning" and "deep, layered learning".

[vii] Even if the activations at the last hidden layer may represent similar features, their location across neurons can shift.

[viii] Excluding perhaps pitched offsets, depending on interpretation

[ix] If each module is trained separately, there will however be no gradient flowing through such connections.

## 8. REFERENCES

Aberdour, M. (2007). Achieving quality in open-source software. *IEEE software, 24*(1).

Alpaydin, E., & Kaynak, C. (1998). Cascading classifiers. *Kybernetika, 34*(4), 369-374.

Basil, V. R., & Turner, A. J. (1975). Iterative enhancement: A practical technique for software development. *IEEE Transactions on Software Engineering*(4), 390-396.

Bengio, Y., Courville, A., & Vincent, P. (2013). Representation learning: A review and new perspectives. *IEEE transactions on pattern analysis and machine intelligence, 35*(8), 1798-1828.

Bittner, R. M., Salamon, J., Bosch, J. J., & Bello, J. P. (2017). *Pitch contours as a mid-level representation for music informatics.* Paper presented at the 3rd AES International Conference on Semantic Audio 2017.

Böck, S., Krebs, F., & Widmer, G. (2016). *Joint Beat and Downbeat Tracking with Recurrent Neural Networks.* Paper presented at the ISMIR.

Boulanger-Lewandowski, N., Bengio, Y., & Vincent, P. (2012). *Modeling temporal dependencies in high-dimensional sequences: application to polyphonic music generation and transcription.* Paper presented at the Proc. of the 29th Int. Conf. on Machine Learning.

Boulanger-Lewandowski, N., Bengio, Y., & Vincent, P. (2013). *Audio Chord Recognition with Recurrent Neural Networks.* Paper presented at the ISMIR.

Breiman, L. (1984). Classification and regression trees.

Chan, T.-S., Yeh, T.-C., Fan, Z.-C., Chen, H.-W., Su, L., Yang, Y.-H., & Jang, R. (2015). *Vocal activity informed singing voice separation with the iKala dataset.* Paper presented at the Acoustics, Speech and Signal Processing (ICASSP).

Choi, K., Fazekas, G., & Sandler, M. (2016). Towards playlist generation algorithms using rnns trained on within-track transitions. *arXiv preprint arXiv:1606.02096*.

Choi, K., Fazekas, G., Sandler, M., & Cho, K. (2017). *Transfer learning for music classification and regression tasks.* Paper presented at the ISMIR.

Chu, H., Urtasun, R., & Fidler, S. (2016). Song from PI: A musically plausible network for pop music generation. *arXiv preprint arXiv:1611.03477*.

Doncieux, S., & Mouret, J.-B. (2014). Beyond black-box optimization: a review of selective pressures for evolutionary robotics. *Evolutionary Intelligence, 7*(2), 71-93.

Dressler, K. (2017). *Automatic transcription of the melody from polyphonic music.* (Doctoral dissertation), Technische Universität Ilmenau, Fakultät für Elektrotechnik und Informationstechnik,

Duarte, M., Oliveira, S., & Christensen, A. L. (2012). *Hierarchical evolution of robotic controllers for complex tasks.* Paper presented at the IEEE Int. Conf. on Development and Learning and Epigenetic Robotics (ICDL)

Durand, S., Bello, J. P., David, B., & Richard, G. (2015). *Downbeat tracking with multiple features and deep neural networks.* Paper presented at the IEEE Int. Conf. on Acoustics, Speech and Signal Processing (ICASSP).

Eerola, T., & Toiviainen, P. (2004). *MIR In Matlab: The MIDI Toolbox.* Paper presented at the ISMIR.

Ellis, H. C. (1965). The transfer of learning.

Elowsson, A. (2016). *Beat tracking with a cepstroid invariant neural network.* Paper presented at the ISMIR.

Elowsson, A. (2018). Polyphonic Pitch Tracking with Deep Layered Learning. *arXiv preprint arXiv:1804.02918*.

Elowsson, A., & Friberg, A. (2015). Modeling the perception of tempo. *The Journal of the Acoustical Society of America, 137*(6), 3163-3177.

Estes, W. K. (1970). *Learning theory and mental development*: Academic Press.

Flood, G. (2016). *Deep learning with a dag structure for segmentation and classification of prostate cancer.* (M. theses), Lund University,

Friberg, A., Schoonderwaldt, E., Hedblad, A., Fabiani, M., & Elowsson, A. (2014). Using listener-based perceptual features as intermediate representations in music information retrieval. *Journal of the Ac. Soc. of Am. (JASA), 136*(4), 1951-1963.

Friedman, J. H. (1991). Multivariate adaptive regression splines. *The annals of statistics*, 1-67.

Gkiokas, A., Katsouros, V., Carayannis, G., & Stajylakis, T. (2012). *Music tempo estimation and beat tracking by applying source separation and metrical relations.* Paper presented at the IEEE Int. Conf. on Ac., Speech and Signal Proc. (ICASSP).

Goodfellow, I., Bengio, Y., & Courville, A. (2016). *Deep learning* (Vol. 1): MIT press Cambridge.

Goodfellow, I., Bulatov, Y., Ibarz, J., Arnoud, S., & Shet, V. (2014). *Multi-digit number recognition from street view imagery using deep convolutional neural networks.* Paper presented at the Int. Conf. on Learning Representations.

Goodfellow, I., Lee, H., Le, Q. V., Saxe, A., & Ng, A. Y. (2009). *Measuring invariances in deep networks.* Paper presented at the Advances in neural information processing systems.

Goto, M., & Muraoka, Y. (1997). *Real-time rhythm tracking for drumless audio signals–chord change detection for musical decisions.* Paper presented at the Working Notes of the IJCAI-97 Workshop on Computational Auditory Scene Analysis.

Gülçehre, Ç., & Bengio, Y. (2016). Knowledge matters: Importance of prior information for optimization. *The Journal of Machine Learning Research, 17*(1), 226-257.

Guo, H., & Gelfand, S. B. (1992). Classification trees with neural network feature extraction. *IEEE Transactions on Neural Networks, 3*(6), 923-933.

Haggblade, M., Hong, Y., & Kao, K. Music genre classification.

Hamel, P., Davies, M., Yoshii, K., & Goto, M. (2013). *Transfer learning in MIR: Sharing learned latent representations for music audio classification and similarity*. Paper presented at the ISMIR.

Hamel, P., & Eck, D. (2010). *Learning Features from Music Audio with Deep Belief Networks.* Paper presented at the ISMIR.

Hawthorne, C., Elsen, E., Song, J., Roberts, A., Simon, I., Raffel, C., . . . Eck, D. (2017). Onsets and Frames: Dual-Objective Piano Transcription. *arXiv preprint arXiv:1710.11153*.

Hawthorne, C., Stasyuk, A., Roberts, A., Simon, I., Huang, C.-Z. A., Dieleman, S., . . . Eck, D. (2018). Enabling factorized piano music modeling and generation with the MAESTRO dataset. *arXiv preprint arXiv:1810.12247*.

He, K., Zhang, X., Ren, S., & Sun, J. (2016). *Deep residual learning for image recognition.* Paper presented at the IEEE conference on computer vision and pattern recognition.

Hinton, G. E., Osindero, S., & Teh, Y.-W. (2006). A fast learning algorithm for deep belief nets. *Neural computation, 18*(7), 1527-1554.

Humphrey, E. J., Bello, J. P., & LeCun, Y. (2012). *Moving Beyond Feature Design: Deep Architectures and Automatic Feature Learning in Music Informatics.* Paper presented at the ISMIR.

Humphrey, E. J., Cho, T., & Bello, J. P. (2012). *Learning a robust tonnetz-space transform for automatic chord recognition.* Paper presented at the Acoustics, Speech and Signal Processing (ICASSP).

Humphrey, E. J., Glennon, A. P., & Bello, J. P. (2011). *Non-linear semantic embedding for organizing large instrument sample libraries.* Paper presented at the Int. Conf. on Machine Learning and Applications (ICMLA).

Hung, Y.-N., & Yang, Y.-H. (2018). *Frame-level instrument recognition by timbre and pitch.* Paper presented at the ISMIR.

Ikemiya, Y., Yoshii, K., & Itoyama, K. (2015). *Singing voice analysis and editing based on mutually dependent F0 estimation and source separation.* Paper presented at the Acoustics, Speech and Signal Processing (ICASSP).

Jansson, A., Humphrey, E., Montecchio, N., Bittner, R., Kumar, A., & Weyde, T. (2017). *Singing voice separation with deep U-Net convolutional networks.* Paper presented at the ISMIR.

Jo, J., & Bengio, Y. (2017). Measuring the tendency of CNNs to learn surface statistical regularities. *arXiv preprint arXiv:1711.11561*.

Joder, C., Essid, S., & Richard, G. (2013). Learning optimal features for polyphonic audio-to-score alignment. *IEEE Transactions on Audio, Speech, and Language Processing, 21*(10), 2118-2128.

Jordan, M. I., & Jacobs, R. A. (1994). Hierarchical mixtures of experts and the EM algorithm. *Neural computation, 6*(2), 181-214.

Juslin, P. N., & Sloboda, J. A. (2001). *Music and emotion: Theory and research*: Oxford University Press.

Kalman, B. L., & Kwasny, S. C. (1997). High performance training of feedforward and simple recurrent networks. *Neurocomputing, 14*(1), 63-83.

Korzeniowski, F., Böck, S., & Widmer, G. (2014). *Probabilistic Extraction of Beat Positions from a Beat Activation Function.* Paper presented at the ISMIR.

Krebs, F., Böck, S., Dorfer, M., & Widmer, G. (2016). *Downbeat Tracking Using Beat Synchronous Features with Recurrent Neural Networks.* Paper presented at the ISMIR.

Kwon, T., Jeong, D., & Nam, J. (2017). *Audio-to-score alignment of piano music using RNN-based automatic music transcription*. Paper presented at the Sound and Music Computing conference (SMC).

Lartillot, O., Toiviainen, P., & Eerola, T. (2008). A matlab toolbox for music information retrieval. In *Data analysis, machine learning and applications* (pp. 261-268): Springer.

LeCun, Y., Bengio, Y., & Hinton, G. (2015). Deep learning. *nature, 521*(7553), 436.

LeCun, Y., Boser, B., Denker, J. S., Henderson, D., Howard, R. E., Hubbard, W., & Jackel, L. D. (1989). Backpropagation applied to handwritten zip code recognition. *Neural computation, 1*(4), 541-551.

Lee, G. K., & Cole, R. E. (2000). The Linux kernel development as a model of open source knowledge creation. *unpub. MS, Haas School of Business, UC Berkeley*.

Li, T. L., Chan, A. B., & Chun, A. (2010). *Automatic musical pattern feature extraction using convolutional neural network.* Paper presented at the Proc. Int. Conf. Data Mining and Applications.

Liang, D., Zhan, M., & Ellis, D. P. (2015). *Content-Aware Collaborative Music Recommendation Using Pre-trained Neural Networks.* Paper presented at the ISMIR.

Lindsay, P. H., & Norman, D. A. (2013). *Human information processing: An introduction to psychology*: Academic press.

Manzelli, R., Thakkar, V., Siahkamari, A., & Kulis, B. (2018). *Conditioning deep generative raw audio models for structured automatic music.* Paper presented at the ISMIR, Paris.

Marolt, M. (2004). A connectionist approach to automatic transcription of polyphonic piano music. *IEEE Transactions on Multimedia, 6*(3), 439-449.

Mauch, M. (2010). *Automatic Chord Transcription from Audio Using Computational Models of Musical Context.* (Ph.D dissertation),

McFee, B., Raffel, C., Liang, D., Ellis, D. P., McVicar, M., Battenberg, E., & Nieto, O. (2015). *librosa: Audio and music signal analysis in python.* Paper presented at the 14th python in science conference.

Miron, M., Carabias-Orti, J. J., & Janer, J. (2014). *Audio-to-score Alignment at the Note Level for Orchestral Recordings.* Paper presented at the ISMIR.

Moon, J., & Sproull, L. (2010). Essence of distributed work. *Online communication and collaboration: A reader*, 125.

Nam, J., Ngiam, J., Lee, H., & Slaney, M. (2011). *A Classification-Based Polyphonic Piano Transcription Approach Using Learned Feature Representations.* Paper presented at the ISMIR.

Ono, N., Miyamoto, K., Kameoka, H., Le Roux, J., Uchiyama, Y., Tsunoo, E., . . . Sagayama, S. (2010). Harmonic and percussive sound separation and its application to MIR-related tasks. In *Advances in music information retrieval* (pp. 213-236): Springer.

Pan, S. J., & Yang, Q. (2010). A survey on transfer learning. *IEEE Transactions on knowledge and data engineering, 22*(10), 1345-1359.

Papadopoulos, H., & Peeters, G. (2008). *Simultaneous estimation of chord progression and downbeats from an audio file.* Paper presented at the IEEE Int. Conf. on Acoustics, Speech and Signal Processing, (ICASSP).

Papadopoulos, H., & Peeters, G. (2011). Joint estimation of chords and downbeats from an audio signal. *IEEE Transactions on Audio, Speech, and Language Processing, 19*(1), 138-152.

Parncutt, R., Sloboda, J. A., Clarke, E. F., Raekallio, M., & Desain, P. (1997). An ergonomic model of keyboard fingering for melodic fragments. *Music Perception: An Interdisciplinary Journal, 14*(4), 341-382.

Peeters, G., & Papadopoulos, H. (2011). Simultaneous beat and downbeat-tracking using a probabilistic framework: Theory and large-scale evaluation. *IEEE Transactions on Audio, Speech, and Language Processing, 19*(6), 1754-1769.

Peretz, I., & Coltheart, M. (2003). Modularity of music processing. *Nature neuroscience, 6*(7), 688.

Perkins, D. N., & Salomon, G. (1992). Transfer of learning. *International encyclopedia of education, 2*, 6452-6457.

Pesek, M., Leonardis, A., & Marolt, M. (2014). *A Compositional Hierarchical Model for Music Information Retrieval.* Paper presented at the ISMIR.

Poliner, G. E., & Ellis, D. P. (2006). A discriminative model for polyphonic piano transcription. *EURASIP Journal on Advances in Signal Processing, 2007*(1).

Quinlan, J. R. (1986). Induction of decision trees. *Machine learning, 1*(1), 81-106.

Read, J., & Hollmén, J. (2014). *A deep interpretation of classifier chains.* Paper presented at the Int. Symp. on Intelligent Data Analysis.

Read, J., Martino, L., & Luengo, D. (2013). *Efficient Monte Carlo optimization for multi-label classifier chains.* Paper presented at the IEEE Int. Conf. on Acoustics, Speech and Signal Processing (ICASSP).

Read, J., Pfahringer, B., Holmes, G., & Frank, E. (2009). *Classifier chains for multi-label classification.* Paper presented at the Joint European Conference on Machine Learning and Knowledge Discovery in Databases.

Rump, H., Miyabe, S., Tsunoo, E., Ono, N., & Sagayama, S. (2010). *Autoregressive MFCC Models for Genre Classification Improved by Harmonic-percussion Separation.* Paper presented at the ISMIR.

Salamon, J., & Gómez, E. (2012). Melody extraction from polyphonic music signals using pitch contour characteristics. *IEEE Transactions on Audio, Speech, and Language Processing, 20*(6), 1759-1770.

Schmidt, E. M., & Kim, Y. (2013). *Learning Rhythm And Melody Features With Deep Belief Networks.* Paper presented at the ISMIR.

Schörkhuber, C., Klapuri, A., Holighaus, N., & Dörfler, M. (2014). *A Matlab toolbox for efficient perfect reconstruction time-frequency transforms with log-frequency resolution.* Paper presented at the AES Int. Conference on Semantic Audio.

Schramm, R., & Benetos, E. (2017). *Automatic transcription of a cappella recordings from multiple singers.* Paper presented at the AES Int. Conference on Semantic Audio.

Schwaber, K. (2004). *Agile project management with Scrum*: Microsoft press.

Sermanet, P., Frome, A., & Real, E. (2014). Attention for fine-grained categorization. *arXiv preprint arXiv:1412.7054*.

Stoller, D., Ewert, S., & Dixon, S. (2018). *Jointly Detecting and Separating Singing Voice: A Multi-Task Approach.* Paper presented at the International Conference on Latent Variable Analysis and Signal Separation.

Sturm, B., Santos, J. F., Ben-Tal, O., & Korshunova, I. (2016). *Music Transcription Modelling and Composition Using Deep Learning.* Paper presented at the 1st Conference on Computer Simulation of Musical Creativity.

Sturm, B. L. (2013). Classification accuracy is not enough. *Journal of Intelligent Information Systems, 41*(3), 371-406.

Sturm, B. L. (2014). A simple method to determine if a music information retrieval system is a "horse". *IEEE Transactions on Multimedia, 16*(6), 1636-1644.

Szegedy, C., Zaremba, W., Sutskever, I., Bruna, J., Erhan, D., Goodfellow, I., & Fergus, R. (2013). Intriguing properties of neural networks. *arXiv preprint arXiv:1312.6199*.

Thickstun, J., Harchaoui, Z., Foster, D. P., & Kakade, S. M. (2018). *Invariances and data augmentation for supervised music transcription.* Paper presented at the 2018 IEEE International Conference on Acoustics, Speech and Signal Processing (ICASSP).

Urzelai, J., Floreano, D., Dorigo, M., & Colombetti, M. (1998). Incremental robot shaping. *Connection Science, 10*(3-4), 341-360.

Valero-Mas, J. J., Benetos, E., & Inesta, J. M. (2016). Classification-based Note Tracking for Automatic Music Transcription.

Van Den Oord, A., Dieleman, S., & Schrauwen, B. (2014). *Transfer learning by supervised pre-training for audio-based music classification.* Paper presented at the ISMIR.

Van Den Oord, A., Dieleman, S., Zen, H., Simonyan, K., Vinyals, O., Graves, A., . . . Kavukcuoglu, K. (2016). *WaveNet: A generative model for raw audio.* Paper presented at the SSW.

vd Boogaart, C. G., & Lienhart, R. (2009). *Note onset detection for the transcription of polyphonic piano music.* Paper presented at the Int. Conf. on Multimedia and Expo (ICME).

Veire, L. V., & De Bie, T. (2018). From raw audio to a seamless mix: creating an automated DJ system for Drum and Bass. *EURASIP Journal on Audio, Speech, and Music Processing, 2018*(1), 13.

Vincent, P., Larochelle, H., Bengio, Y., & Manzagol, P.-A. (2008). *Extracting and composing robust features with denoising autoencoders.* Paper presented at the the 25th Int. Conference on Machine Learning.

Viola, P., & Jones, M. (2001). *Rapid object detection using a boosted cascade of simple features.* Paper presented at the Computer Vision and Pattern Recognition (CVPR).

Wang, Z., Wang, X., & Wang, G. (2018). Learning fine-grained features via a CNN tree for large-scale classification. *Neurocomputing, 275*, 1231-1240.

Weninger, F., Kirst, C., Schuller, B., & Bungartz, H.-J. (2013). *A discriminative approach to polyphonic piano note transcription using supervised non-negative matrix factorization.* Paper presented at the IEEE Int. Conf. on Acoustics, Speech and Signal Processing (ICASSP).

White, C., & Pandiscio, C. (2015). *Do Chord Changes Make Us Hear Downbeats?* The University of Massachusetts, Amherst.

Xiao, T., Xu, Y., Yang, K., Zhang, J., Peng, Y., & Zhang, Z. (2015). *The application of two-level attention models in deep convolutional neural network for fine-grained image classification.* Paper presented at the IEEE Conf. on Computer Vision and Pattern Recognition (CVPR).

Yang, L.-C., Chou, S.-Y., & Yang, Y.-H. (2017). *MidiNet: A convolutional generative adversarial network for symbolic-domain music generation.* Paper presented at the ISMIR.

Zenz, V., & Rauber, A. (2007). *Automatic chord detection incorporating beat and key detection.* Paper presented at the IEEE Int. Conf. on Signal Processing and Communications (ICSPC).

Zhang, N., Donahue, J., Girshick, R., & Darrell, T. (2014). *Part-based R-CNNs for fine-grained category detection.* Paper presented at the European Conf. on computer vision.

Zhao, B., Feng, J., Wu, X., & Yan, S. (2017). A survey on deep learning-based fine-grained object classification and semantic segmentation. *Int. Journal of Automation and Computing, 14*(2), 119-135.